\documentclass[a4paper, 10pt, conference]{ieeeconf}

\IEEEoverridecommandlockouts
\usepackage[english]{babel}

\usepackage{amssymb}
\usepackage{amsmath}
\usepackage{hyperref}
\usepackage[capitalise]{cleveref}
\usepackage{url}
\usepackage{graphicx}
\usepackage{latexsym}
\usepackage{algorithm}
\usepackage{algpseudocode}
\usepackage[binary-units]{siunitx}

\usepackage{todonotes}

\usepackage{subfig}

\usepackage{listings}

\usepackage[acronym]{glossaries}
\glsdisablehyper
\newacronym{fd}{FD}{Fall Detection}
\newacronym{afd}{AFD}{Automatic Fall Detection}
\newacronym{adl}{ADL}{Activities of Daily Living}
\newacronym{fds}{FDS}{Fall Detection System}
\newacronym{fdt}{FDT}{Fall Detection Technique}
\newacronym{har}{HAR}{Human Activity Recognition}
\newacronym{rnn}{RNN}{Recurrent Neural Network}
\newacronym{lstm}{LSTM}{Long Short-Term Memory}
\newacronym{imu}{IMU}{Inertial Measurement Unit}
\newacronym{knn}{KNN}{K-Nearest Neighbors}
\newacronym{ann}{ANN}{Artificial Neural Network}
\newacronym{rbf}{RBF}{Radial Basis Function}
\newacronym{ppca}{PPCA}{Probabilistic Principal Component Analysis}
\newacronym{lda}{LDA}{Linear Discriminant Analysis}
\newacronym{svm}{SVM}{Support Vector Machine}
\newacronym{ml}{ML}{Machine Learning}
\newacronym{dl}{DL}{Deep Learning}
\newacronym{fsm}{FSM}{Finite State Machine}

\title{\LARGE \bf
Online Fall Detection using Recurrent Neural Networks
}

\author{Mirto Musci, Daniele De Martini, Nicola Blago, Tullio Facchinetti and Marco Piastra\\
Dept. of Electrical, Computer and Biomedical Engineering\\Universit\`a degli Studi di Pavia, via Ferrata, 5 -- 27100 Pavia, Italy\\
	mirto.musci@unipv.it, \{daniele.demartini01, nicola.blago01\}@ateneopv.it,\\\{tullio.facchinetti, marco.piastra\}@unipv.it
}

\begin{document}

\maketitle
\thispagestyle{empty}
\pagestyle{empty}

\begin{abstract}
Unintentional falls can cause severe injuries and even death, especially if no immediate assistance is given.
The aim of \glspl{fds} is to detect an occurring fall.
This information can be used to trigger the necessary assistance in case of injury.
This can be done by using either ambient-based sensors, e.g.~cameras, or wearable devices. 

The aim of this work is to study the technical aspects of \glspl{fds} based on wearable devices and artificial intelligence techniques, in particular \gls{dl}, to implement an effective algorithm for on-line fall detection.
The proposed classifier is based on a \gls{rnn} model with underlying \gls{lstm} blocks.
The method is tested on the publicly available SisFall dataset, with extended annotation, and compared with the results obtained by the SisFall authors.
\end{abstract}

\glsresetall

\section{INTRODUCTION}

Unintentional falls are the leading cause of fatal injury and the most common cause of nonfatal trauma-related hospital admissions among older adults.
As stated in~\cite{reportOnFalls}, more than 25\% of people aged over 65 years old falls every year increasing to 32\%--42\% for those over 70.
Moreover, 30\%--50\% of people living in long-term care institutions fall each year, with almost half of them experiencing recurrent falls.
Falls lead to 20\%--30\% of mild to severe injuries and 40\% of all injury deaths.
The average cost of a single hospitalization for fall-related injuries in 65 years old people reached \$17483 in the US in 2004, with a forecast to \$240 billion of total costs by 2040.

Elderly people are not the only group that is heavily affected by unintentional falls: any person with some sort of fragility is part of similar statistics.
Examples include any kind of mild disability and post-operative patients.
The situation is worsened when people live alone, so they may not receive immediate assistance in case of accident~\cite{Fleming2227}.

The aim of \glspl{fds} is to promptly detect occurring falls in real-time and hence insuring a remote notification, so that timely aid can be given.
\Gls{fd} can be a non-trivial issue, indeed, while a human being can easily recognise a fall when the falling subject is in sight, it might be difficult for a machine, since a fall can only be detected indirectly.

In the light of these premises, the aim of this paper is to study the technical aspects of using wearable devices for \glspl{fds} and \gls{dl} techniques to develop an effective fall detection algorithm.
In particular, our goal is to be able to detect falls on-line and in real time, directly on the sensor stream, with a relatively simple architecture.
The proposed solution is compared against a baseline based on statistical analysis.

The paper is organized as follows.
Firstly \cref{s:state_of_the_art} discusses some related works, then \Cref{s:proposed_method} describes in detail the proposed method, while \cref{s:implementation_and_results} shows the experimental result achieved on a selected dataset.
Finally, \cref{s:conclusions} concludes the paper.

\section{STATE OF THE ART}
\label{s:state_of_the_art}

The analysis of the state of the art is divided into two parts.
The first part encompasses a list of relevant detection techniques.
In the second part, given the goal of using a neural network approach, available datasets are revised with the objective of selecting the most adequate one.

\subsection{Detection techniques}

\Glspl{fds} can be classified into two main categories: ambient-based and based on wearable devices~\cite{mohamed2014}.
Ambient-based sensors are mainly based on video cameras, either standard or RGB-D~\cite{Baldewijns2012017,Kepski2014}.
These techniques are intrusive in terms of privacy, and they adapt poorly to the case of highly mobile persons who are not restricted to confined areas.
Moreover, \gls{afd} in video streams still represents a difficult problem to address even for recent \gls{dl} methods, such as in~\cite{feng2014deep}.

Wearable devices, on the other hand, offer portability as they can be used regardless of the user location.
The most widely adopted sensor to equip wearable devices is the \mbox{3-axis} accelerometer due to its low cost and tiny size.
The common availability of accelerometers in smartphones opens the possibility to use such devices as a cost-effective sensory device.
Beside being widespread and economically affordable, smartphones provide a robust and powerful hardware platform (i.e., processor, screen and radio), which allows to implement fully self-contained monitoring applications.

Rotational sensors (gyroscopes) are also used in some works~\cite{Bou08}.
The work proves the relevance of information provided from gyros for \gls{fd}, since the authors achieve very good results with the use of that type of measurements alone.
However, many commercial smartphones are not equipped with gyros, making this approach not viable on these devices.
In \cite{Li09} the authors use both accelerometers and gyros targeting fast response and low computational requirements.

\Gls{afd} from wearable sensor data is currently an open problem, with multiple approaches in the literature.
The common procedure is to record the raw acceleration sensor readings, to filter them and then to apply a feature extraction method that classifies activities as falls or other activities, the so-called \gls{adl}.
In~\cite{projectGravity,sisfall}, the occurrence of a fall is detected using a threshold technique that relies on statistics on quantities such as the acceleration magnitude and its standard deviation.
In~\cite{tFall}, instead, the \gls{fds} is considered as an anomaly detection system and the authors draw a comparison between two different \gls{ml} techniques: \gls{knn} and \gls{svm}.
A simple feed-forward \gls{ann} is used in \cite{abbate2012}, which is fed with a set of features obtained in a time window of three seconds.
In cascade of that, a \gls{fsm} is used to discard the false positive matches.

In~\cite{gibson2016}, the authors combine different techniques to enhance the prediction of the classifier.
Indeed, they investigate the use of \gls{ann}, \gls{knn}, \gls{rbf}, \gls{ppca} and \gls{lda}.
The authors of~\cite{Ordonez2016} utilize a \gls{dl} technique to detect falls.
The architecture is composed by two parts: a Convolutional Network and a \gls{rnn}.
The former is leveraged to automatically extract features from sensor signals, while the latter defines a temporal relationship between the samples.
Finally, in~\cite{tsinganos2017} a \gls{fsm}-like model is used to detect the fall episodes.
Such episodes are then decomposed into features that are sent to a \gls{knn} classifier to distinguish between falls and \glspl{adl}.

\subsection{Datasets}

Since the goal of this work is to use a \gls{dl} approach, selecting the appropriate dataset for training and validation of the model represents a crucial issue.
The basic requirements imposed on the dataset are the presence of both falls and \glspl{adl}, and the availability of detailed measurements from onboard sensors.
For this purpose, $6$ datasets were considered for the application of the machine learning approach \cite{DLR, tFall, projectGravity, mobifall, sisfall, UMAFall}.
The datasets are listed in \cref{tab:datasets}, along with the number of different subjects involved in the experiments, and the number of different recorded activities.

\begin{table}
\centering
\begin{tabular}{lc|ccc}
\textbf{Dataset} & \textbf{Ref.} & \textbf{Number of} & \textbf{Number of} & \textbf{Sensing} \\
		& 		& \textbf{subjects} & \textbf{activities} & \textbf{device}\\
\hline
DLR & \cite{DLR} & 16 & 6 & Smartphone\\
tFall & \cite{tFall} & 10 & 8 & Smartphone\\ 
Project Gravity & \cite{projectGravity} & 3 & 19 & Smartphone\\
MobiFall & \cite{mobifall} & 24 & 13 & Smartphone\\
SisFall & \cite{sisfall} & 38 & 34 & Custom\\
UMAFall & \cite{UMAFall} & 17 & 11 & Custom\\
\end{tabular}
\caption{List of available datasets that have been considered for the proposed \gls{dl} approach.\label{tab:datasets}}
\end{table}

For the purpose of this work, the SisFall dataset was selected since it is deemed the most complete among the ones considered.
In fact, it includes the largest amount of data, both in terms of number and heterogeneity of \gls{adl} and subjects.
Moreover, the protocol is validated by a medical staff and each action to be performed is described in a video clip recorded with an instructor.

In addition, most of the other listed datasets obtained the data from smartphones, while SisFall adopts a dedicated custom measuring device that is fixed on body as a belt buckle.
Such device included two different models of 3D accelerometers and a gyroscope.
Sensors data were sampled at a frequency of $200~\si{\hertz}$.

\section{PROPOSED SOLUTION}
\label{s:proposed_method}

The aim of this work is to detect occurring falls with temporal precision and accuracy.
The adopted method is based on \gls{ml} and, in particular, on a \gls{dl} approach.
Therefore, the objective requires specific provisions both in dataset preparation and training.

\subsection{Dataset and Labeling}

The training of a detection system based on a dataset of measured values requires specific \emph{labeling} that includes temporal annotations, i.e. the explicit indication of temporal intervals associated to the relevant events of interest.
For this purposes, we considered three classes of events:
\begin{itemize}
	\item \textbf{FALL}: this class identifies the time interval when the person is experiencing a state transition that leads to a catastrophic change of state, i.e., a fall.
	\item \textbf{ALERT}: the time interval in which the person is in a dangerous state transition; this state may lead to a fall, or the subject may be able to avoid the fall.
	\item \textbf{BKG}: the default time interval when the person is in control of his/her own state.
\end{itemize}

Since we are interested in detecting falls, the BKG class is considered to absorb all the activities that are not related to a fall (\glspl{adl}), such as walking, jumping, walking up the stairs, sitting on a chair, and so on.
All these activities are included in the considered dataset.
Our classification focuses in particular on catastrophic transitions (i.e.~falls).

On the other hand, we need datasets in which all data sequences are annotated by temporal intervals that correspond to FALL and ALERT events, while every other interval is considered as BKG by default.
It is worth to outline that this temporal labeling is not available in any datasets in \cref{tab:datasets}.
Therefore, the existing data belonging to the SisFall dataset was manually labeled to enable the use as a training and validation set for the neural network.

\subsection{Deep Learning Approach}
\label{s:dl}

\begin{figure}
	\centering
	\includegraphics[width=0.55\columnwidth]{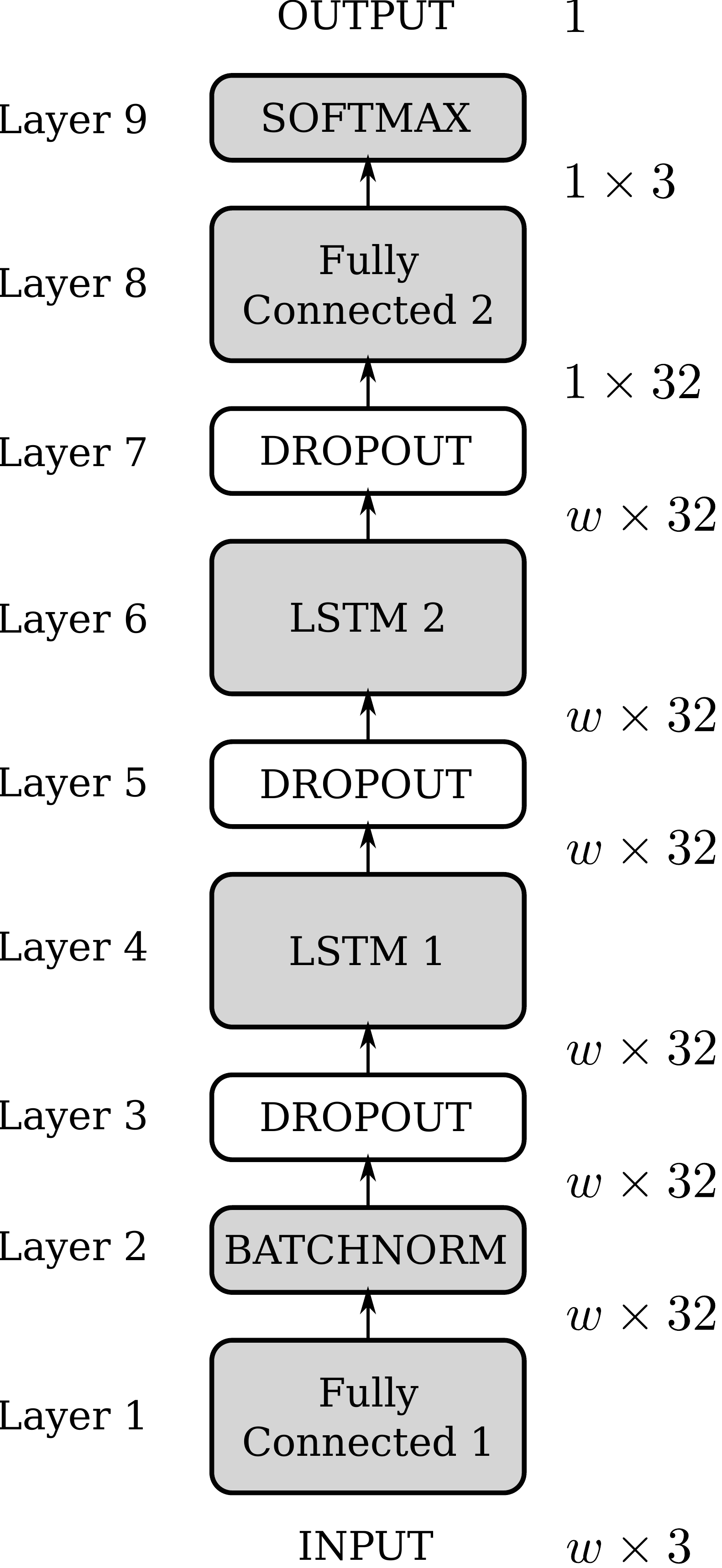}
	\caption{The model architecture of the proposed solution.
	White blocks are active during the training phase only and are removed when doing inference on the device. The input size depends on the number of selected features (in our case the 3 axis of the accelerometer) and on the size $w$ of the windows to be fed into the model. The output size depends on the number of classes (i.e. BKG, ALERT and FALL).\label{fig:architecture}}
\end{figure}

One of the goals of the proposed application is to run on top of relatively cheap and resource-constrained devices, such as a microprocessor-equipped \textit{smart sensor}, an embedded platform or a smartphone.
Therefore, the selection of the \gls{dl} architecture that implements the \textit{inference} module, i.e. the component that performs the actual detection, must take into account this requirement.
As a result, we propose a relatively simple network architecture that has been adapted from~\cite{chevalier}.

The core of the \gls{dl} architecture is depicted in \cref{fig:architecture}.
The network is based on two \gls{lstm} cells~\cite{lstm} connected in series (Layers~4 and~6).
Each cell has a inner dimension of $32$ units.
The input preprocessing is performed by the fully connected Layer~1, while a second fully connected layer (Layer~8) collects the output from the \gls{lstm} cell at Layer~6 and feeds its output to the final SoftMax Layer (Layer~9) that provides the classification in the three classes described above.

With respect to the network structure proposed in~\cite{chevalier}, the adopted solution includes a batch normalization layer~\cite{batchnorm} (Layer~2) to regularize input data, and three dropout layers~\cite{dropout} (Layers~3, 5 and~7).
The latter are used during network training to improve generalization, while they are removed in the deployed inference module.

\subsection{Training and Inference}

The training of \gls{lstm} cells is based on the idea of temporal unfolding~\cite{lstm}.
Using the temporal unfolding, the input is processed in sub-sequences, called \textit{windows}, having predefined length $w$.
Each \gls{lstm} cell is cascaded into exactly $w$ copies of itself.
Each copy receives the output of the previous cell of the cascade and the input from the window at the corresponding index.

Temporal unfolding allows the training of a \gls{rnn}, such as the \gls{lstm}, as if it was a non-recurrent deep network.
This approach entails that input data sequences in both the training set and the live sensor readings must be arranged in windows having the same fixed size $w$.

As will be seen in \cref{s:implementation}, the window size $w$ represents a hyperparameter which tuning is critical for the effectiveness of the whole network architecture.

\section{IMPLEMENTATION AND RESULTS}
\label{s:implementation_and_results}

This section briefly describes the implementation details of the proposed solution and shows the results obtained in comparison to the statistical classifier proposed in~\cite{sisfall}.

\subsection{Labeling Procedure}

As discussed above, none of the considered datasets contains the detailed temporal annotation that is required for the purpose of applying the \gls{dl} classification method based on the network described in \cref{s:dl}.
This fact is true for the SisFall dataset as well.
As a result, further annotation work was performed to make the dataset suitable for our purposes.

The performed annotation procedure consisted in adding temporal intervals to the existing data sequences in order to mark the time ranges corresponding to FALL and ALERT classes, while the remaining intervals are implicitly associated to the BKG class.
To simplify the classification, the BKG class is also assigned to the immediate aftermath of each FALL, namely when the person is experiencing several secondary minor state transitions like bumping or rolling.

The labeling procedure has been carried out by the authors using a software tool developed on purpose.
The SisFall dataset is provided with a set of video clips describing the actions corresponding to each class of movement in the SisFall dataset.
However, these video clips only describe the prototypical action performed by the instructor, not each of the actions performed by SisFall volunteers.
Therefore, we elicited criteria of our own for temporal annotation by comparing the aforementioned video clips with the corresponding sensor data recordings.

\begin{figure}
	\centering
	\includegraphics[width=0.8\columnwidth]{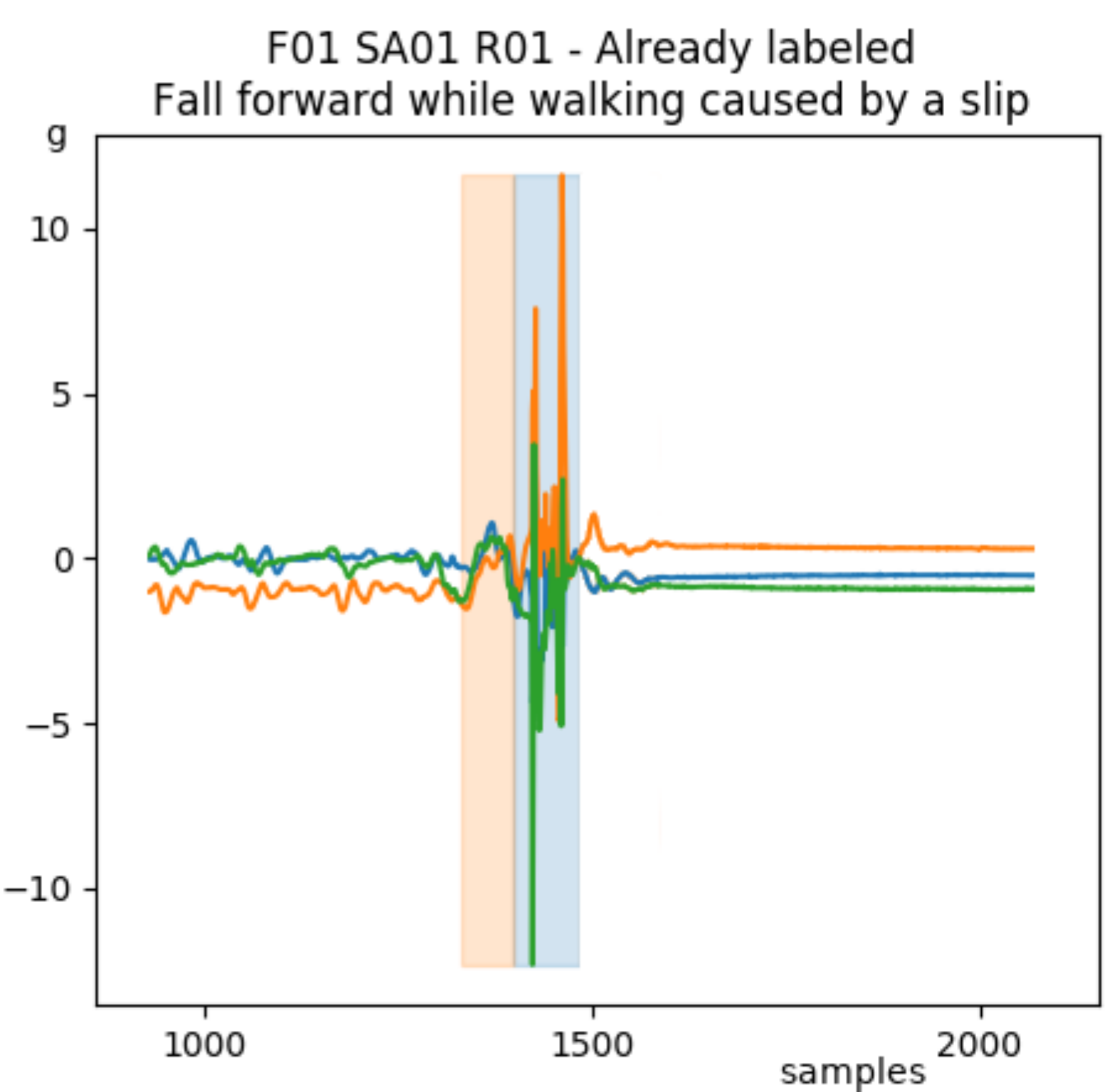}
	\caption{
Example of temporally-annotated data.
        The orange and pale blue areas indicate a ALERT and a FALL class respectively, while the remainder of the sequence is classified as BKG by default.\label{fig:labeled}}
\end{figure}

\Cref{fig:labeled} shows an example of temporally-annotated data sequence from SisFall.
In this example the three solid lines describe the readings of the 3D accelerometer for a maneuver in which a walking person falls down due to a slip.
The initial slip is classified as an ALERT (in orange), followed by the FALL proper (in pale blue).
As described before, the aftermath of the FALL is ignored.

%
%
The result of the described labeling procedure is openly available online\footnote{\url{https://bitbucket.org/unipv_cvmlab/sisfalltemporallyannotated/}} as an extension to the SisFall dataset.

\subsection{Software Implementation and Training}
\label{s:implementation}

The architecture described in \cref{s:proposed_method} has been implemented using the TensorFlow library~\cite{tensorflow}, with the Python programming language.
All training and testing procedures were performed on a Dell 5810 workstation, equipped with a Nvidia Quadro K5000 GPU.

To train the model, all the annotated SisFall sequences were divided into windows having width $w$ and a partial overlapping induced by a \textit{stride} of length $s$.
In general, each window can span over a time frame that may contain a set of samples belonging to different classes.
To translate the temporal (i.e. per-sample) labels described above into window labels we adopted to the following criteria:

\begin{itemize}
	\item each window containing at least 10\% of readings labeled as FALL was tagged as FALL altogether;
	\item each non-FALL window in which the majority of samples was labeled as ALERT is tagged as ALERT;
	\item any other window was tagged to the BKG class.
\end{itemize}

A first problem to be addressed while training is that, for any reasonable choice of window width $w$, the three classes above are not  equally represented in the preprocessed SisFall dataset.
In other words, the number of windows labeled as BKG is much larger than the number of windows labeled as FALL or ALERT.
In fact, in SisFall sequences, BKG sequences may last for several seconds, whereas a single FALL will last for a couple of seconds at most, and in many cases it is as short as $500~\si{\milli\second}$.
This aspect causes a strong imbalance in the training dataset, in which the BKG windows can be as much as $50$ times more represented than FALLs.

\begin{figure}
	\centering
	\includegraphics[width=0.90\columnwidth]{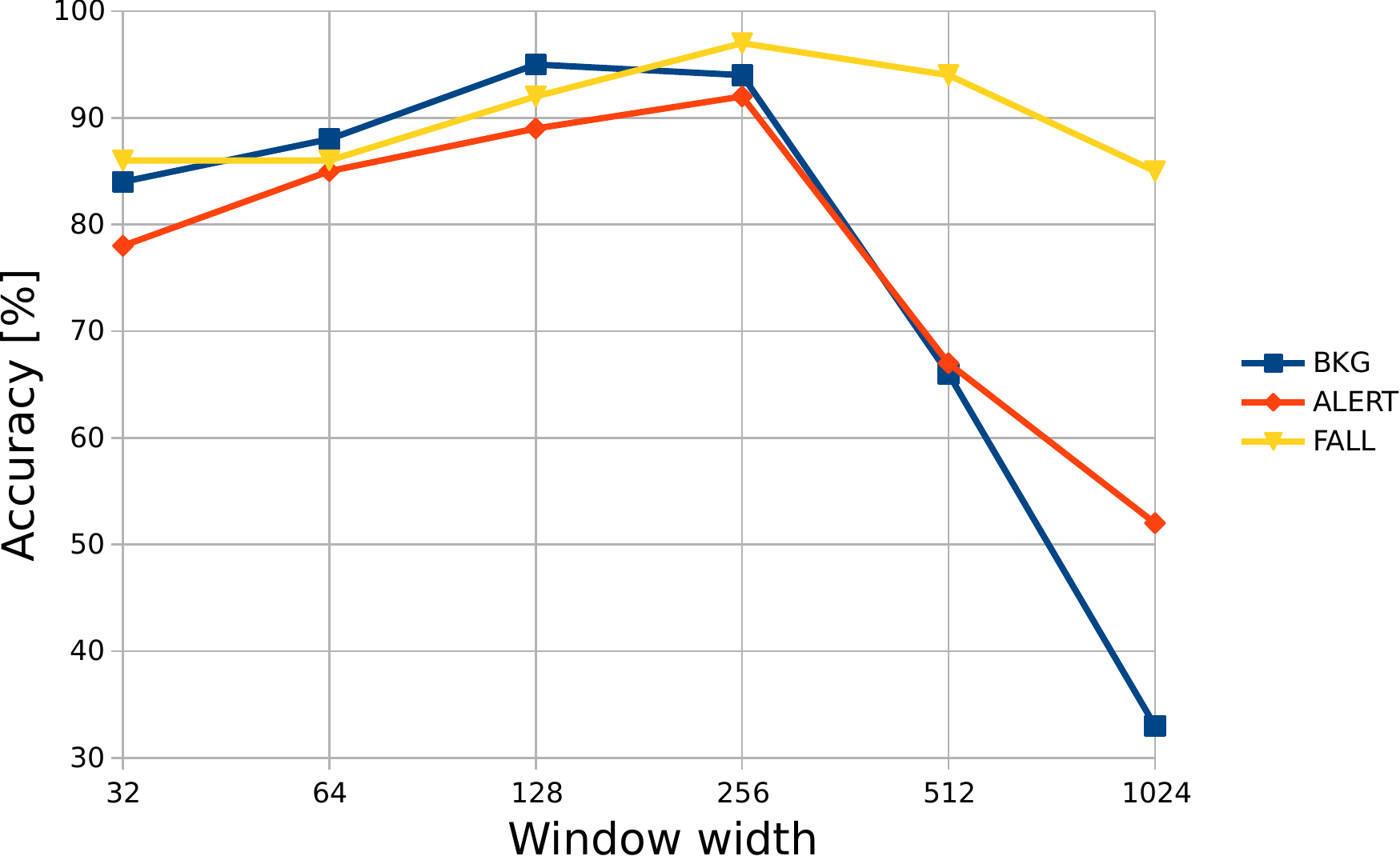}
	\caption{Accuracy of the classifier for different window widths $w$ and a stride of 50\% of $w$, for the three classes and using a balancing loss function.\label{fig:accuracy}}
\end{figure}

\begin{figure}
	\centering
	\subfloat[]{\includegraphics[width=0.78\columnwidth]{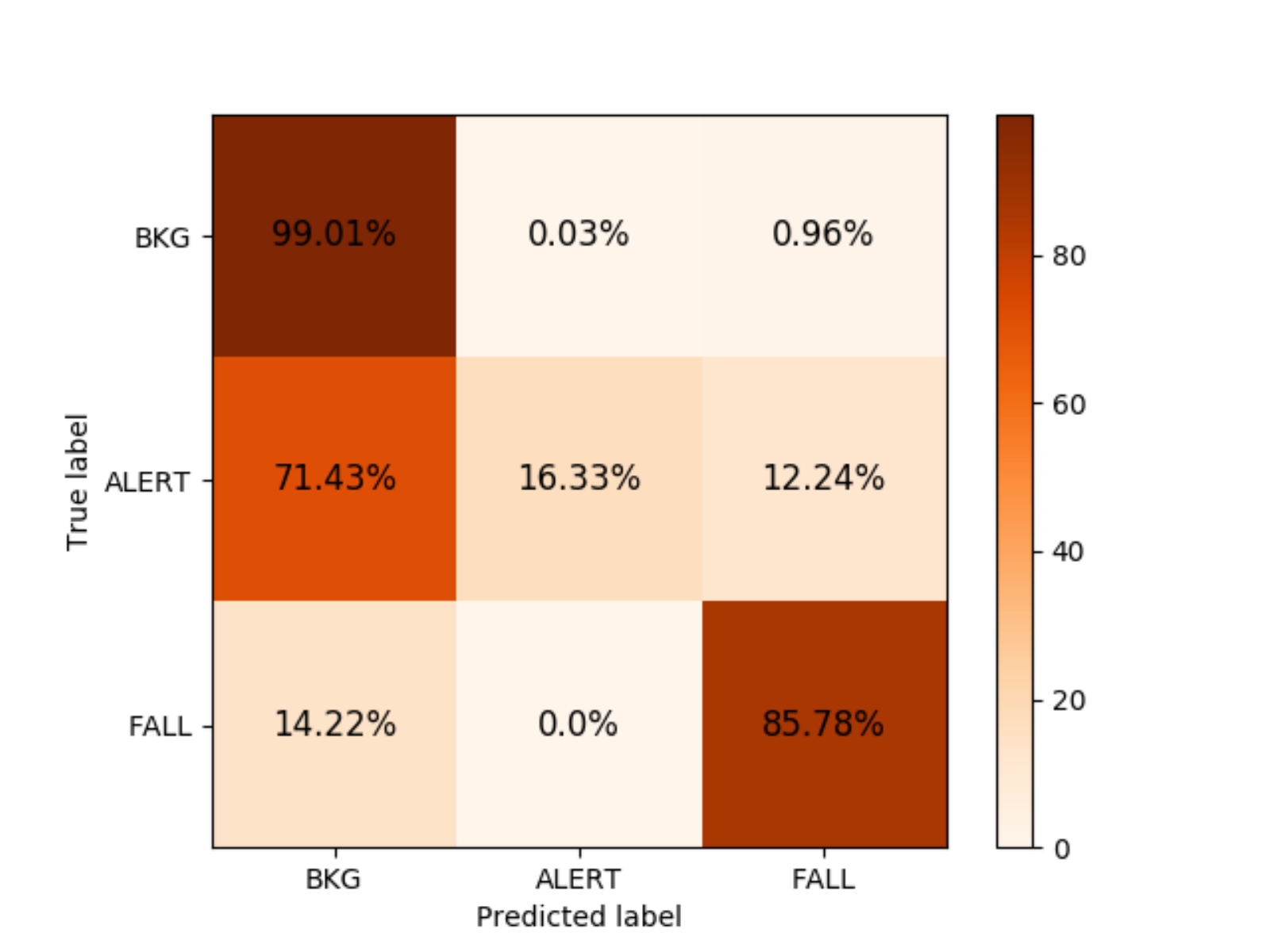}\label{fig:confusion_mat_old}}\\
	\subfloat[]{\includegraphics[width=0.78\columnwidth]{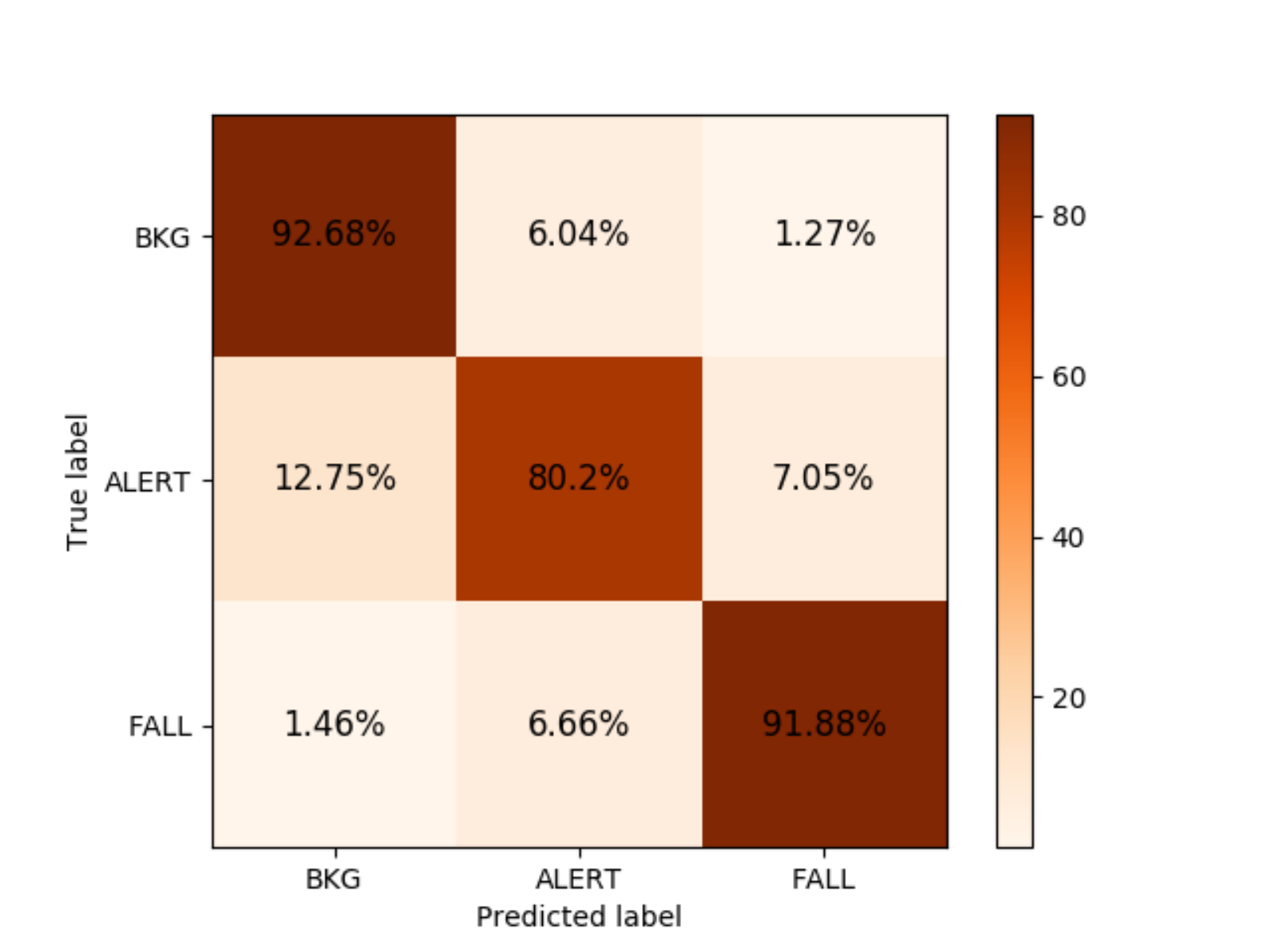}\label{fig:confusion_mat_new}}\\
	\subfloat[]{\includegraphics[width=0.78\columnwidth]{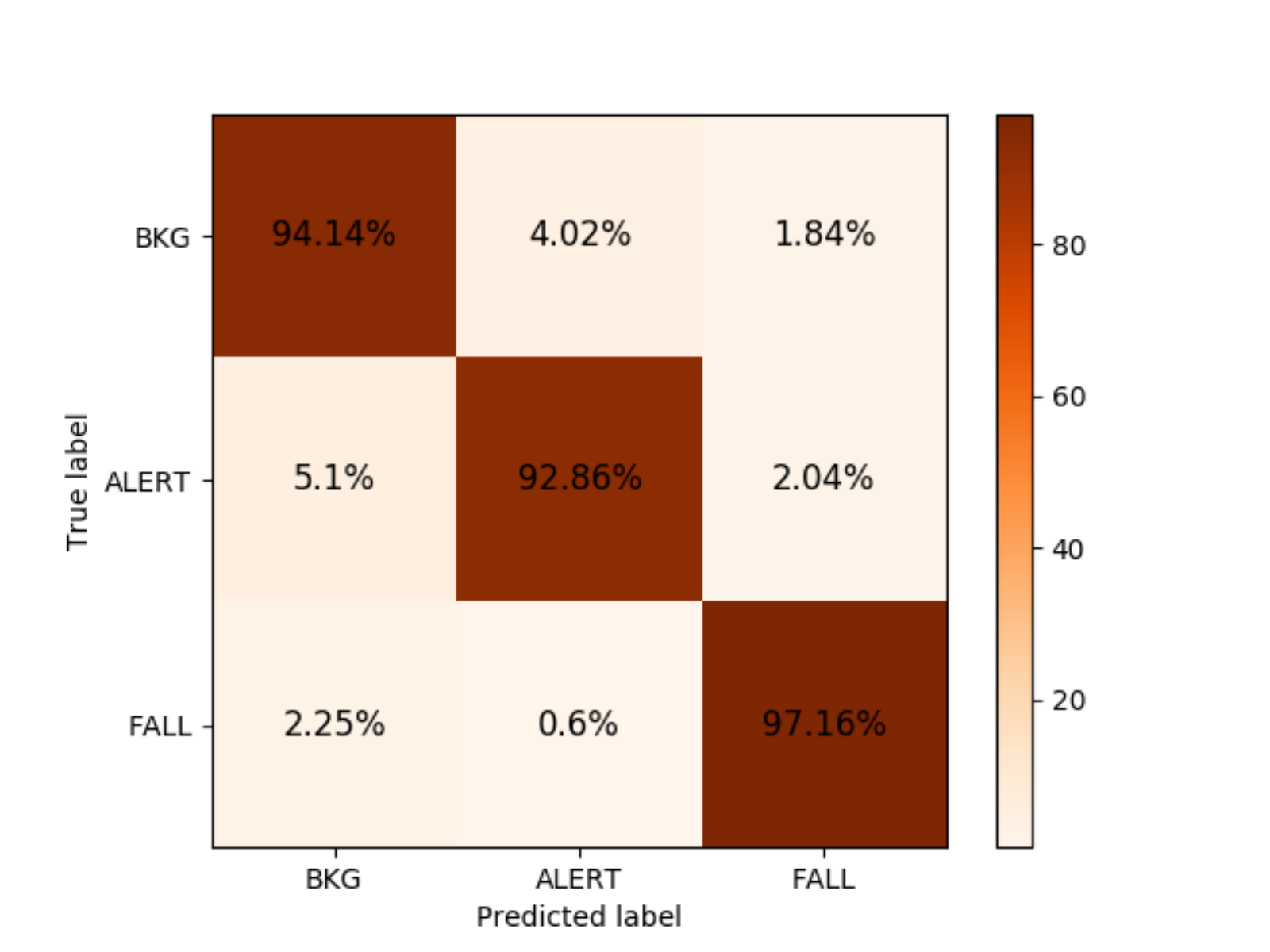}\label{fig:confusion_mat_w256_s128_vote}}
	\caption{Confusion matrices obtained by using the baseline cross-entropy loss function (case (a)) and with the custom weighted loss function (b) with $w = 128$. Figure (c) shows the confusion matrix using the optimal window width $w = 256$ with the weighted loss function after a thorough hyperparameters optimization.\label{fig:confusion_matrices}}
\end{figure}

\Cref{fig:confusion_mat_old} shows the confusion matrix resulting from the training with a standard loss function and $w = 128$.
As it can be expected, BKG activities are classified very well, whereas FALLs are poorly detected and ALERTs are almost left undetected.

Our proposed solution to such unbalancing is to define a balancing loss function.
In particular, we implemented a weighted cross-entropy, where each sample (a window in our case) contributes to the gradient descent with a multiplier factor that depends on the inverse frequency of its class members in the training datasets.

More formally, being $N$ the set of all training windows, we define $B$, $A$ and $F$ as the subsets of $N$ belonging to the BKG, ALERT and FALL classes, respectively. The basic multiplier $m_i$ for each sample $i$ is defined as:
\begin{equation}
m_i=
\begin{cases}
	1 & \forall i \, \in B \\
	\left\vert{B}\right\vert/\left\vert{A}\right\vert &
	\forall i \,\in A\\
	\left\vert{B}\right\vert/\left\vert{F}\right\vert &
	\forall i \,\in F\\

\end{cases}
\end{equation}

and thus, denoting $\mathcal{L}_i$ the value of the standard cross-entropy loss function computed on the $i$-th sample, we define the weighted loss function as:
\begin{equation}
	\mathcal{L}_{weight}=
	\sum_i^{\left\vert{N}\right\vert}
	m_{i}\mathcal{L}_i
\end{equation}
\Cref{fig:confusion_mat_new} shows the confusion matrix resulting from training with $\mathcal{L}_{weight}$, with a consequent drastic rise of the accuracy, above all for the ALARM class.

As anticipated, the window width $w$ is a critical hyperparameter, while the choice of the stride $s$ is more dictated by practical considerations: a very low value of $s$ entails a substantial increase in the computational burden.
The choice of the most effective values for hyperparameters $w$ and $s$ was made via a sensitivity analysis, in which window widths were varied between $32$ and $1024$ and stride values corresponding to 25\%, 50\% or 75\% of $w$ were considered.

\Cref{fig:accuracy} shows the values of accuracy related to all the three classes for each window width value and a fixed stride of 50\%.
As it can be seen, a window width of 256, corresponding to $1.28~\si{\second}$, and a stride of 50\% resulted to be the most effective in \gls{fd}.

As it can be seen in the figure, the classification errors were drastically reduced.
Finally, \cref{fig:confusion_mat_w256_s128_vote} shows the results after an extensive grid-search optimization of every \gls{dl} hyperparameter such as number of epochs, batch size and learning rate, using the optimal window width $w = 256$.

\subsection{Comparison with the Baseline}

In the original paper that introduced the SisFall dataset~\cite{sisfall}, several statistical indicators were proposed for \gls{fd}.
Such indicators were applied to entire sequences and a classification was assigned if, at some point in time, the selected indicator was above a predefined threshold.
In~\cite{sisfall}, it is shown that the so-called $C_8$ and $C_9$ classifiers, presented below, were the most effective in the classification approach used in the paper.

Let us denote with $\sigma^2(a_\mathit{ax}^k)$ the variance of the acceleration along the $\mathit{ax}$ axis in the $k$-th window, where $\mathit{ax} \in \{x, y, z \}$ is one of the $3$ axes of the 3D accelerometer.
The $C_8$ and $C_9$ indicators are thus defined by \cref{eq:c8,eq:c9}, respectively.

\begin{equation}
	\label{eq:c8}
	C_8 := \sqrt{\sigma^2(a_x^k) + \sigma^2(a_z^k)}
\end{equation}
\begin{equation}
	\label{eq:c9}
	C_9 := \sqrt{\sigma^2(a_x^k) + \sigma^2(a_y^k) + \sigma^2(a_z^k)}
\end{equation}

The difference between $C_8$ and $C_9$ is that the former relies on the standard orientation of the acquisition device used in SisFall: such device was tied to the belt of the performer with the $y$ axis along the vertical direction, pointing below, and the $z$ axis facing the forward direction.
Clearly, $C_9$ is a more realistic classifier for situations in which the above assumptions do not hold, e.g.~for a smartphone in a different carry position and it has been chosen as the comparison baseline.

To present a meaningful comparison between our proposed classification method and the $C_9$ indicator, we applied the latter to ech window of size $w$ and we defined suitable thresholds for both ALERT and FALL classes.

For the latter purpose we applied an iterative technique -- on the same training dataset used for \gls{dl} -- to derive the best thresholds (in terms of accuracy) for the detection of both FALL and ALERT classes.

\begin{figure}
	\centering
	\includegraphics[width=0.78\columnwidth]{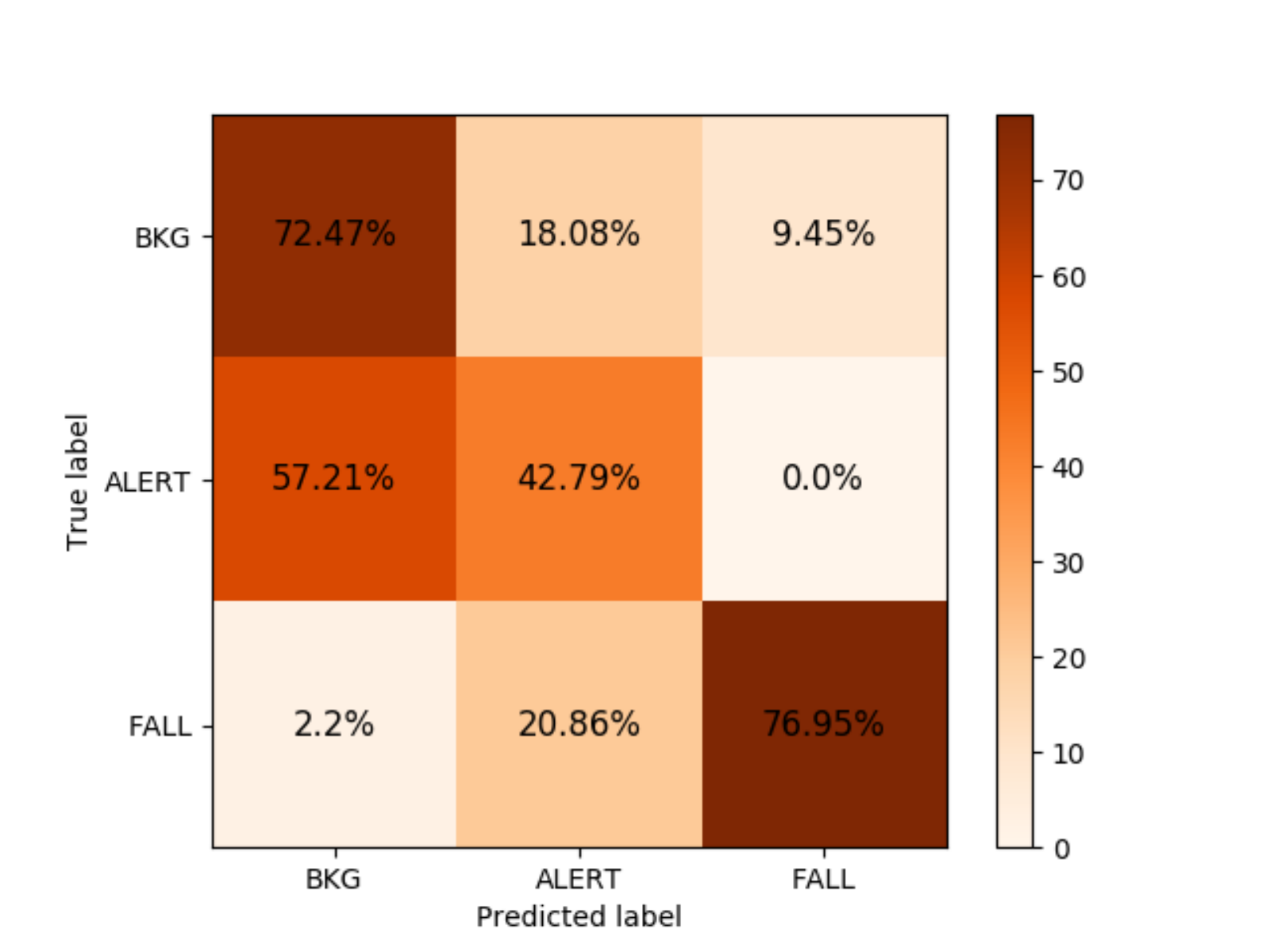}
	\caption{Confusion matrix obtained with the $C_9$ classifier with $w=256$ and stride of 50\%.
	\label{fig:confusion_c9_vote}}
\end{figure}

\Cref{fig:confusion_c9_vote} shows the confusion matrix obtained with the $C_9$ classifier applied to windows of width $256$ and stride 50\%.
It is clear from a simple visual inspection that the proposed solution significantly outperforms the $C_9$ indicator.

\begin{figure}
	\centering
	\includegraphics[width=0.85\columnwidth]{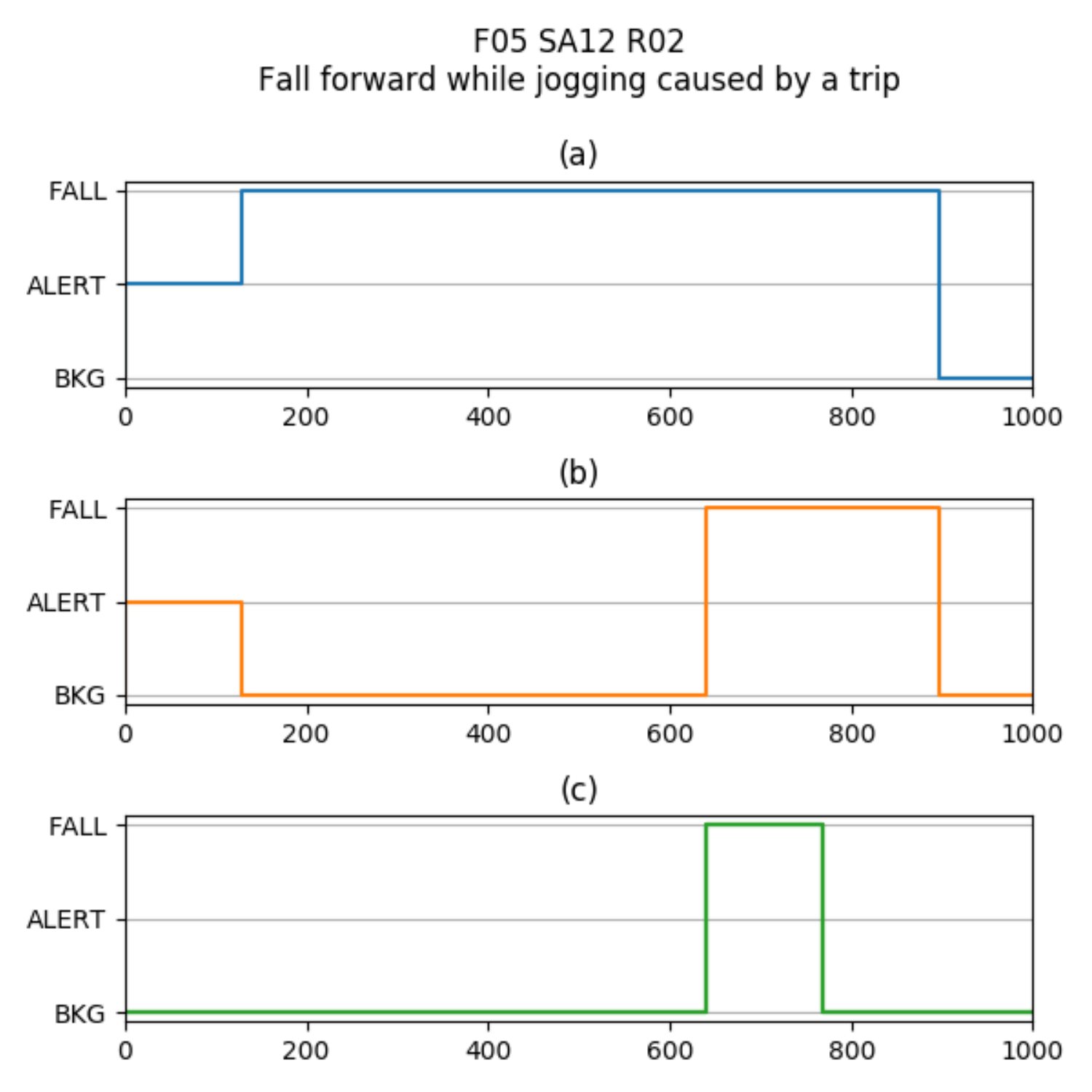}
	\caption{Classification behavior of the $C_9$ indicator (case (a)) and the proposed \gls{dl} approach (b), with respect to ground-truth labeling (c) for falling as a result of jogging.\label{fig:comparison}}
\end{figure}

\Cref{fig:comparison} shows the behavior of both the baseline and our proposed \gls{dl} model when compared to the manually assigned \textit{ground-truth} labels in a challenging scenario.

Indeed, high-dynamic activities, such as jogging, are hardly distinguishable from falls for the baseline.
Our proposal reaches a better overall classification, even if the initial transition from a standing still into jogging produces the incorrect classification of the first window as an ALERT.
We can also notice that our proposal tends to capture the aftermath of a fall as part of the FALL state.

\section{CONCLUSIONS}
\label{s:conclusions}

The purpose of the work was to apply \gls{dl} techniques to the \gls{afd} problem.
More specifically, \gls{rnn} based on \gls{lstm} blocks were used as the classifier.

After an attentive review of available datasets, the choice fell upon the SisFall dataset, which was deemed the most appropriate for the task at hand.
The assumed requiremen of real-time, online detection led to reconsider the labeling method used in SisFall and to carry out a new and more accurate labeling work, suitable to be used by the \gls{dl}-based approach.
In turn, the introduction of temporal labeling caused a substantial imbalance of classes in the dataset.
This issue was solved by introducing a weighted loss function during the training phase.

The proposed solution achieved a high precision on fall detection, outperforming the results obtained by the SisFall $C_9$ indicator.

\bibliographystyle{IEEEtran}
\bibliography{biblio}

\end{document}